\documentstyle [bezier,epsfig,12pt,preprint,tighten,aps]{revtex}
\begin{document}

\draft

\begin{titlepage}
\rightline{July 2000}
\rightline{Revised September 2000}
\vskip 2cm
\centerline{\large \bf  
Is the $\nu_\mu \to \nu_{s}$ oscillation solution to the
atmospheric neutrino anomaly}
\centerline{\large \bf
excluded by the superKamiokande data?}
\vskip 1.1cm
\centerline{R. Foot\footnote{Email address:
Foot@physics.unimelb.edu.au}}
\vskip .7cm
\centerline{{\it School of Physics}}
\centerline{{\it Research Centre for High Energy Physics}}
\centerline{{\it The University of Melbourne}}
\centerline{{\it Victoria 3010 Australia}}
\vskip 2cm

\centerline{Abstract}
Recently the SuperKamiokande collaboration have claimed
that their data exclude the $\nu_\mu \to \nu_s$ solution
to the atmospheric neutrino anomaly at more than 99\% C.L.
We critically examine this claim.

\vskip 1cm
\noindent

\end{titlepage}
Something mysterious with neutrinos is a foot.
It is clear that about
half of the upward going atmospheric $\nu_\mu 's$
are mising\cite{sk,atmos}. Furthermore, about half of the solar
$\nu_e 's$
have also disappeared\cite{solar}.
There is also strong evidence that 
$\nu_e \leftrightarrow \nu_\mu$ oscillations take
place with small mixing angles from the LSND experiment\cite{lsnd}. 
An elegant explanation of these facts is that
each neutrino oscillates maximally with an approximately
sterile partner, with small angles between generations\cite{fv}.
For the status of the maximal $\nu_e \to \nu_s$ solution
to the solar neutrino problem, see Ref.\cite{smok}.
The status of the maximal $\nu_\mu \to \nu_s$ solution to
the atmospheric neutrino problem is the subject of this paper.

As was pointed out sometime ago\cite{2},
both $\nu_\mu \to \nu_s$ and $\nu_\mu \to \nu_\tau$
oscillations are able to explain the sub-GeV and multi-GeV
superKamiokande single ring events (while 2 flavour
$\nu_\mu \to \nu_e$ oscillations cannot because there
is no observed anomaly with the electron events\cite{fvy}).
Recently, however, the SuperKamiokande Collaboration have argued that 
the $\nu_\mu \to \nu_s$ oscillation explanation of
the observed deficit of
atmospheric neutrinos is disfavoured at 
more than 99\% C.L.\cite{3}, while the interpretation
interms of $\nu_\mu \to \nu_\tau$ oscillations fits 
all of their data extremely well. 
This conclusion relies on an analysis
of the upward through going muon data (UTM), the partially contained 
events with $E_{visible} > 5\ GeV$(PC) and the neutral current
enriched multi-ring events (NC).
These data sets lead to slightly different expectations 
for the $\nu_\mu \to \nu_\tau$ vs $\nu_\mu \to \nu_s$
oscillations because of earth matter effects for
$\nu_\mu \to \nu_s$ oscillations
which are important for UTM\cite{lip} and PC events\cite{lip98,2}
while neutral current interactions in the detector are utilized
for the NC events\cite{hall}.
These three data sets, obtained from Ref.\cite{3} (for
1100 live days), are shown in Figures 1a,b,c.
Also shown is the theoretically expected result
for maximal $\nu_\mu \to \nu_s$ oscillations
with $\delta m^2 = 3\times 10^{-3}\ eV^2$ also obtained
from Ref.\cite{3}.
SuperKamiokande analyse the data by taking particular
ratios and have not as yet
provided detailed justification of the
systematic uncertainties in the theoretically expected rates. 

Let us discuss each of the three data sets in turn:
\vskip 0.2cm
\noindent
a) Upward through going muons:
The overall normalization of the through going muon
fluxes have an estimated 20 \% uncertainty, however
the uncertainty in the expected shape of the zenith
angle distribution is significantly lower
(for some discussion of these uncertainties, see Ref.\cite{5,6}).
A recent estimate\cite{5} of the uncertainty in the vertical/horizontal
ratio due to the uncertainties in the atmospheric 
fluxes is of order 4\%. This systematic uncertainty
is dominated by the uncertainty in the ratio $K/\pi$ 
produced in the atmosphere from the interactions of cosmic
rays\cite{5}. In addition there will be
other uncertainties in the shape of the zenith
angle distribution due to the uncertainty in 
the energy dependence of the neutrino - nucleon cross section 
\footnote{
Uncertainties in the energy dependence of the  cross section 
leads to uncertainties 
in the expected shape of the zenith angle distribution of UTM events, 
because the zenith angle dependence of
the atmospheric neutrino flux is energy dependent.}
and from cosmic ray muons masquerading as neutrino induced
muons. The latter uncertainty, while mainly affecting the
most horizontal bin ($-0.1 < \cos\Theta < 0$) may be very important, as
we will show.
\vskip 0.2cm
\noindent
b) Partially contained events (with $E_{visible} > 5\ GeV$).
The systematic uncertainty in the expected normalization  
of these events is quite large, again of order 20\%\cite{5,6}. 
The systematic uncertainty on the expected shape of 
the zenith angle distribution
of these events should be relatively small ($\stackrel{<}{\sim}
\ few \%$ in the up/down ratio).
\vskip 0.2cm
\noindent
c) Neutral current enriched multi-ring events.
The systematic uncertainty in the expected normalization  
is again quite large, of order 20-40\% due 
to the highly uncertain cross sections (as well as
the uncertain atmospheric fluxes).
The uncertainty in the expected shape of the 
zenith angle distribution will of course be much smaller,
but may be significant (i.e. of order 5\% in the up/down ratio).
The uncertainty is due in part to the uncertainty in
the relative contributions
due to $\nu_e$ interactions (which are expected to be
approximately up/down
symmetric) and $\nu_\mu$ interactions (which are up/down asymmetric due
to the oscillations affecting the upward going $\nu_\mu$'s).
The relative contributions due to the neutral current weak interactions
and the charged current weak interactions will also be uncertain.
In addition to the cross section uncertainties there are also the
uncertainties in the scattering angle distribution between
the angles of the multi-ring events and the incident neutrino. 

Note that the  normalization uncertainties between the three
data sets will be 
largely uncorrelated because of the different energy
ranges for the atmospheric neutrino fluxes and also because of the
different cross sections involved. 
Nevertheless, some weak correlation between UTM and
PC events may be expected. 
While analysing the data using ratio's does eliminate
the normalization uncertainty, the remaining uncertainties
will be important for the data sets a) and c).
Furthermore, a 
conclusion based on particular ratios could only be
robust if it agreed with a $\chi^2$ fit of the binned data points.
SuperKamiokande are in the best position to do this 
for their data, and we hope that they will do this at some point.

In the meantime we will do this using
the superKamiokande theoretical Monte-Carlo results for their
given test point of maximal mixing with
$\delta m^2 = 3 \times 10^{-3}\
eV^2$ (which is not expected to be the best fit for
$\nu_\mu \to \nu_s $ oscillations).
We define the $\chi^2$ by: 
\begin{equation}
\chi^2_{total} = \chi^2_{UTM} + \chi^2_{PC} + \chi^2_{NC},
\end{equation}
with
\begin{equation}
\chi^2_{y} = \sum^{10}_{i=1} \left({data_y (i) - f_y \times theory_y (i) \over
\delta data_y (i)}\right)^2 + 
\left({f_y - 1\over \delta f}\right)^2,
\label{5}
\end{equation}
where $y = UTM, PC, NC$ and the sum runs over the 10 zenith angle
bins, 
and $\delta data_y$ is the statistical uncertainty in
the data, $data_y(i)$. 
The normalization factor, $f_y$ parameterizes the
overall normalization uncertainty in the
theoretical expected value, $theory_y (i)$, 
and $\delta f_y$ is
the expected normalization uncertainty, and we take
$\delta f_y = 0.2$ for $y=UTM,PC,NC$.
It is understood that  $\chi^2_y$ is minimized with respect
to $f_y$. 

Doing this exercise (using the superKamiokande experimental data,
$data_y (i), \delta data_y (i)$ and also the superKamiokande
theoretically expected results $theory_y(i)$ for maximal mixing with
$\delta m^2 = 3\times 10^{-3}\ eV^2$),
we find that $\chi^2_y$ is minimized when
$f_{UTM} \simeq 0.90$, $f_{PC} \simeq 0.87$ and
$f_{NC} \simeq 1.07$. 
In Figures 2a,b,c we compare the data with $f_y theory_y(i)$,
which is the theoretical prediction for $\delta m^2 = 3\times 
10^{-3} \ eV^2$ (neglecting systematic uncertainties in
the shape).
We obtain the following $\chi^2_y$ values:
\begin{eqnarray}
\chi^2_{UTM} = 17.0 \ \ for \ 10 \ degrees \ of \ freedom,\nonumber \\
\chi^2_{PC} = 13.4 \ \ for \ 10 \ degrees \ of \ freedom,\nonumber \\
\chi^2_{NC} = 16.0 \ \  for \ 10 \ degrees \ of \ freedom.
\end{eqnarray}
Thus we obtain $\chi^2_{total} \simeq 46 $ for $30$ degrees of freedom
which corresponds to an allowed C. L. of about 3\%.
While this allowed C.L. is low, it is only a {\it lower limit}
because we haven't varied $\delta m^2$ or incorporated the systematic
uncertainties in the shape of $theory_y (i)$,
which we now discuss.

Varying $\delta m^2$ within the allowed region identified
from a fit to the contained events should 
improve $\chi^2_{PC}$ somewhat as well as slightly improving
$\chi^2_{UTM,NC}$. 
For example, for $\delta m^2 = 5\times 10^{-3}\ eV^2$ using our code
developed in Ref.\cite{2} we find that $\chi^2_{PC} \simeq 12$ (c.f. 
$13.4$ for $\delta m^2 = 3\times 10^{-3}\ eV^2$).

With regard to the UTM and NC events the effect of 
systematic uncertainties on $\chi^2$ can be very dramatic.
We illustrate this by introducing a slope factor s(i) defined
by
\begin{equation}
s(i) = 0.95 +  0.01*i,
\label{xxu}
\end{equation}
where $i=1,...,10$ (with $i=1$ the vertical upward going bin).
In Eq.(\ref{5}) we replace $f_{UTM}(i) \to s(i)*f_{UTM}(i)$,
which is roughly within 
the estimated 1-sigma systematic uncertainty for
the UTM events.  In fact,
this would be roughly equivalent to reducing the atmospheric $K/\pi$
ratio by about $30-40\%$ to be compared with the estimated
$25\%$ uncertainty for the $K/\pi$ ratio\cite{5}. While the uncertainty
in the $K/\pi$ ratio may be the largest single contribution to the uncertainty
in the shape of the zenith angle distribution of UTM events, 
the total systematic uncertainty in the shape of 
the zenith angle distribution gets many contributions
\footnote{
Due to e.g uncertainty in the energy dependence of the neutrino
nucleon cross section, uncertainty in the interaction length
of the cosmic rays in the upper atmosphere, modelling of
the atmosphere, uncertainty in the primary cosmic ray energy spectrum
and composition of cosmic rays etc.}
which is why the slope factor in Eq.(\ref{xxu}) might be
expected to be roughly  within the 1-sigma systematic uncertainty.
With the above slope factor, we find $\chi^2_{UTM} \simeq 13$, which
represents a significantly improved fit.
From our ealier discussion, the systematic
uncertainties in the shape of the zenith angle distribution of 
the events for UTM and NC events are expected to be completely uncorrelated.
This means that the best fit for the NC events can have a slope factor with
a slope of a different sign, and this is needed to improve the fit.
To illustrate the effect then, for NC we replace
$f_{NC}(i) \to f_{NC}(i)/s(i)$ and find
$\chi^2_{NC} \simeq 12$. 
This demonstrates that a $\chi^2$ fit to the
three data sets incorporating the systematic uncertainties and
varying $\delta m^2$ would be expected to reduce $\chi^2_{total}$ by
at least 9 leading to a $\chi^2_{total}$ of about 37 or less.
This corresponds to an allowed C.L. of 15\% or more. Of course a 
global fit of all the superKamiokande data 
gives a much larger allowed C.L. because of the excellent fit
of the $\nu_\mu \to \nu_s$ oscillations to the lower energy 
contained events (both sub-GeV and multi-GeV)\cite{val}. 
The results obtained for UTM and NC events using the 
slope factor $s(i)$ are given by the dotted lines in Figure 2a,c.

We would also like
to emphasise that the poor $\chi^2$ fit
for UTM events is due largely to the most 
horizontal bin ($-0.1 < \cos\Theta < 0$). 
Excluding this bin we find that
\begin{equation}
\chi^2_{UTM} = 12.5  \ for \ 9 \ bins
\end{equation}
{\it excluding} any systematic uncertainty in the shape
of the zenith angle distribution (i.e. with $s(i)=1$).
The reason for questioning the horizontal bin is clear:
It is expected that the systematic uncertainty for the most horizontal
bin should be relatively large. This is because atmospheric
muons can contribute. (In fact 
the Kamiokande collaboration\cite{kam} made the cut $\cos\Theta < -0.04$
and incorporated large systematic errors for this bin).
SuperKamiokande, in their published analysis of 537 days
\cite{superk} included
the whole horizontal bin, and made an estimate of the contamination
of atmospheric muons in this bin (of order 4\%) and subtracted
it off. This is based on an extrapolation from $\cos\Theta > 0$
where the background falls off exponentially. This exponential
assumption is not discussed in any detail, and needs to
be justified if it can be. In fact, from their
Figure 1\cite{superk}, 
which compares the distribution of through-going muons
near the horizon observed by superKamiokande for regions with thick and
thin rock overburden, it seems possible that the atmospheric
muon background could be higher by a factor of two
or three or even more.
This is rather important. For example, if a background of $10\%$ 
is assumed (which means that we must
lower the superKamiokande data value by $6\%$ for this bin), 
then we obtain a $\chi^2_{UTM} \simeq 14$
for 10 degrees of freedom (excluding the effects
of the systematic uncertainties in the shape of
the zenith angle distribution, i.e. $s(i) = 1$)
or $\chi^2_{UTM} \simeq 11$ including the modest slope factor in
Eq.(\ref{xxu}).
Unless the level of 
contamination of atmospheric muons in the horizontal
bin can be rigorously justified, 
it is probably safest to exclude the horizontal
bin altogether because the systematic uncertainties may
be so large as to make it too uncertain to be useful\footnote{
In terms of analysis with ratio's we suggest that the vertical
be defined as $-1 < \cos\Theta < -0.5$ and the 
horizontal as $-0.5 < \cos\Theta < -0.9$.}.

Thus, we have shown that a $\chi^2$
analysis of the recent upward through going muon binned data, partially
contained events with $E_{visible} > 5\ GeV$ and
neutral current enriched multi-ring events
does not exclude maximal $\nu_\mu \to \nu_s$
oscillation solution to the atmospheric neutrino
problem with any significant confidence level. 
This is not inconflict with the superKamiokande 
results since they fit three particular ratio's rather
than the binned data. However it does show that 
the conclusion that the $\nu_\mu \to \nu_s$ osillations
are disfavoured does depend on how one analyses the data.
Furthermore, the overall fit (i.e. including also 
the lower energy single ring events) of the $\nu_\mu \to
\nu_s$ oscillations to the superKamiokande data
is good. Fortunately 
future data will eventually decide the issue.
In the meantime, important work needs to be
done on carefully estimating and checking the possible
systematic uncertainties.

\vskip 0.4cm
\noindent
{\bf Acknowledgement}
\vskip 0.4cm
\noindent
The author would like to thank Paolo Lipari and Ray Volkas
for many related discussions/communications and for
comments on a preliminary version of the paper.
The author is an Australian Research Fellow.

\vskip 0.6cm
\noindent
{\bf Figure Captions}
\vskip 0.4cm
\noindent
Figure 1: SuperKamiokande data for the upward through going muons
(Fig.1a), partially contained events with $E_{visible} > 5\ GeV$
(Fig.1b) and neutral current enriched multi-ring events
(Fig.1c), all obtained from Ref.\cite{3}. Also shown are the 
superKamiokande expected results for
maximal $\nu_\mu \to \nu_s$ oscillations with
$\delta m^2 = 3 \times 10^{-3}\ eV^2$, also obtained from
Ref.\cite{3}.
\vskip 0.3cm
\noindent
Figure 2: Same as Figure 1 except that the theoretical
expectation for
maximal $\nu_\mu \to \nu_s$ oscillations with
$\delta m^2 = 3 \times 10^{-3}\ eV^2$, are renormalized by
an overall scale factor (as discussed in the text).
In Figures 2a and 2c,
the dotted line includes the effect of a modest correction
to the expected shape of the 
zenith angle distribution given by Eq.(\ref{xxu}), 
as discussed in the text.

\newpage
\epsfig{file=f1a.eps,width=15cm}
\newpage
\epsfig{file=f1b.eps,width=15cm}
\newpage
\epsfig{file=f1c.eps,width=15cm}
\newpage
\epsfig{file=f2a.eps,width=15cm}
\newpage
\epsfig{file=f2b.eps,width=15cm}
\newpage
\epsfig{file=f2c.eps,width=15cm}
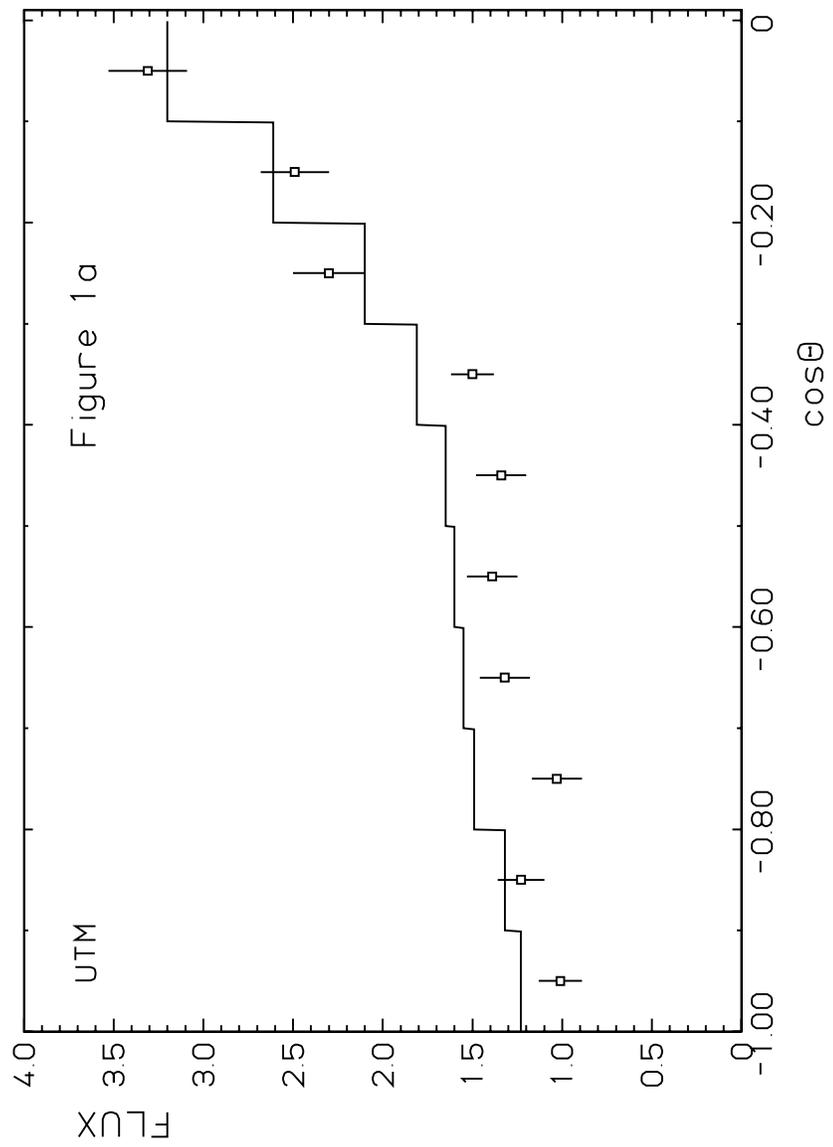
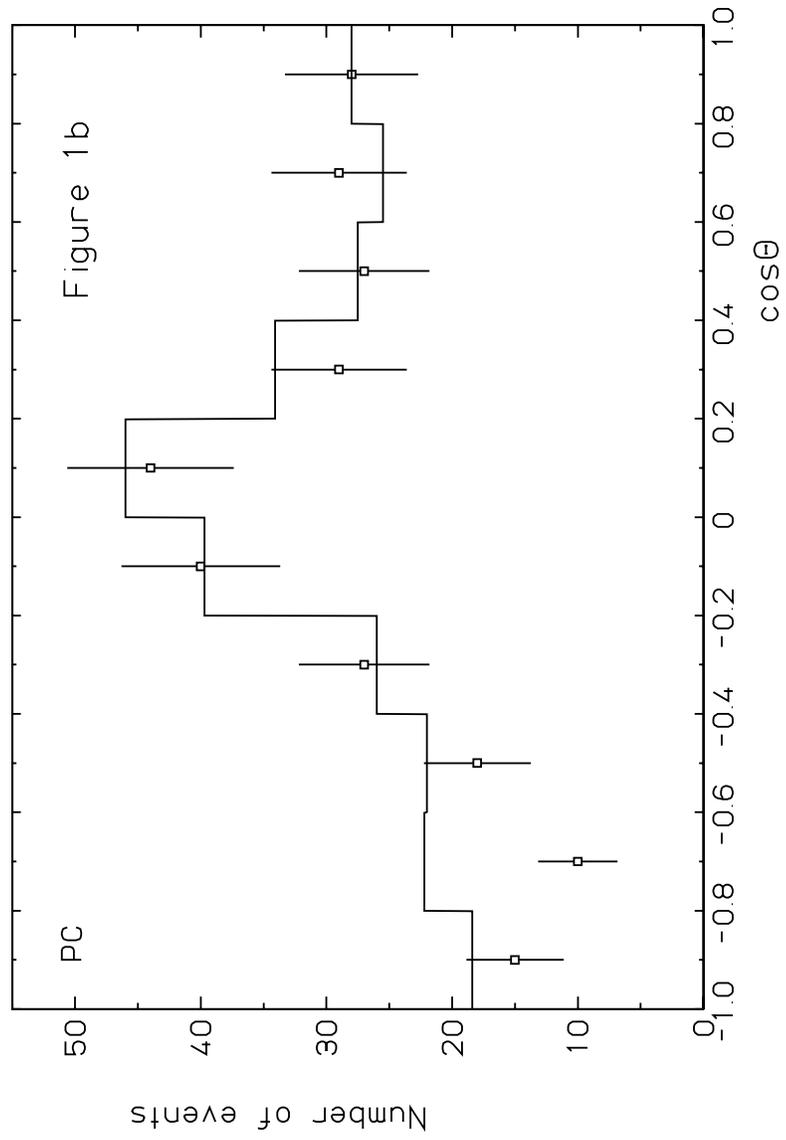
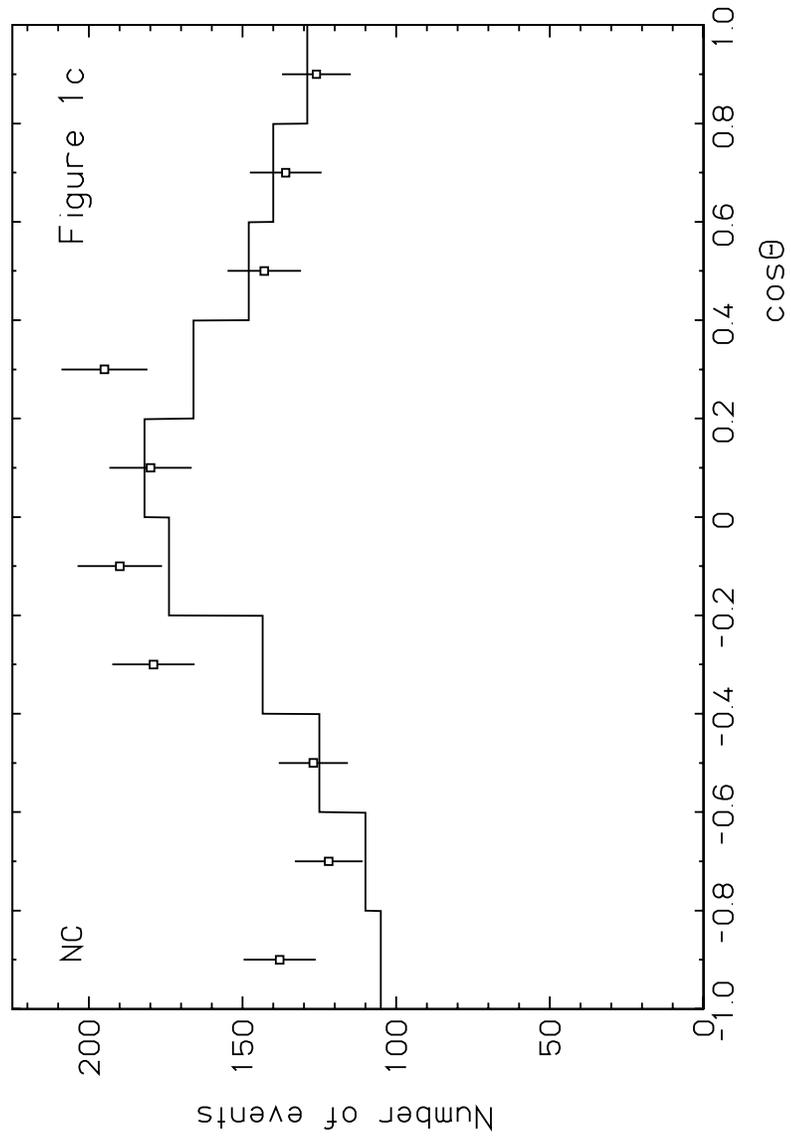
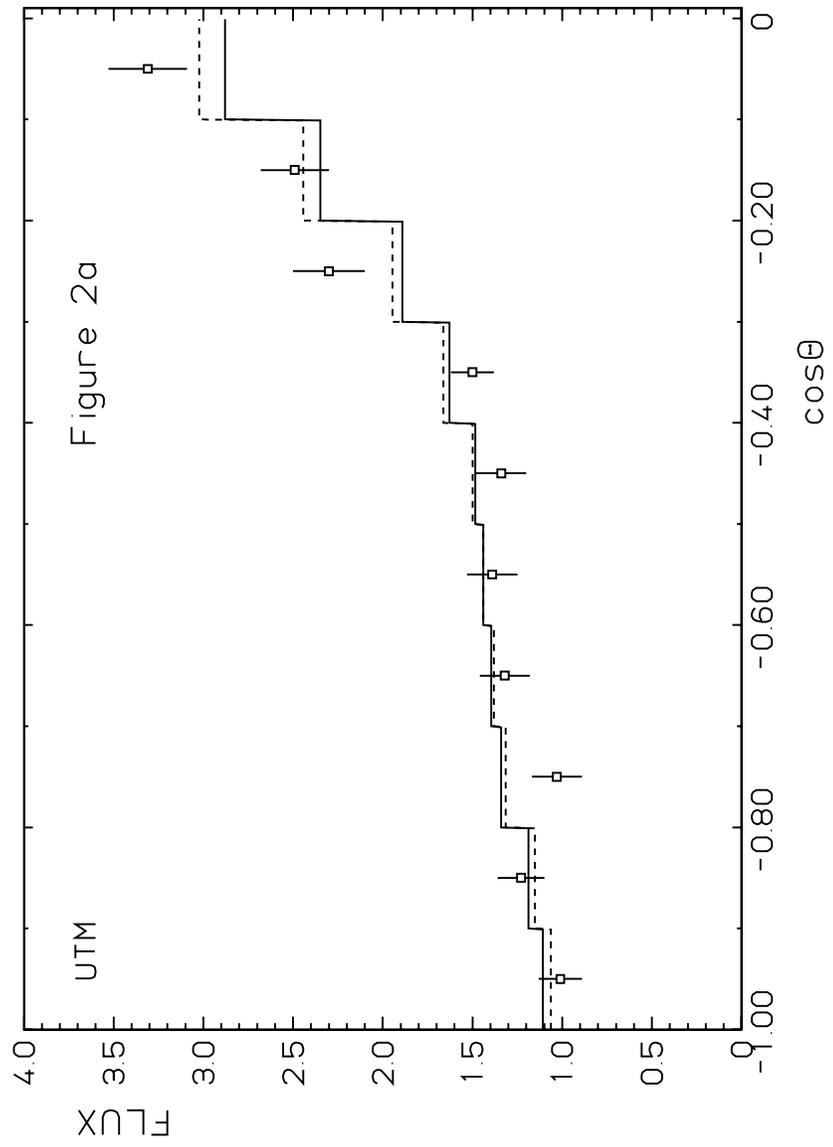
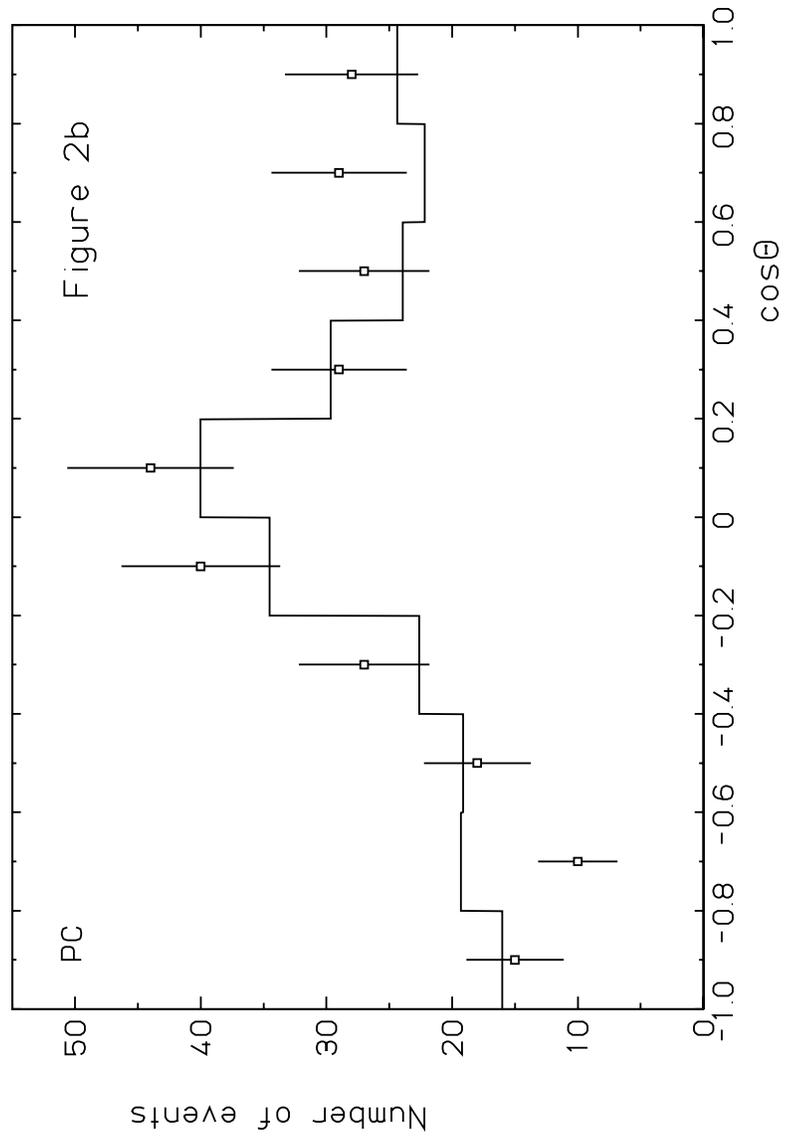
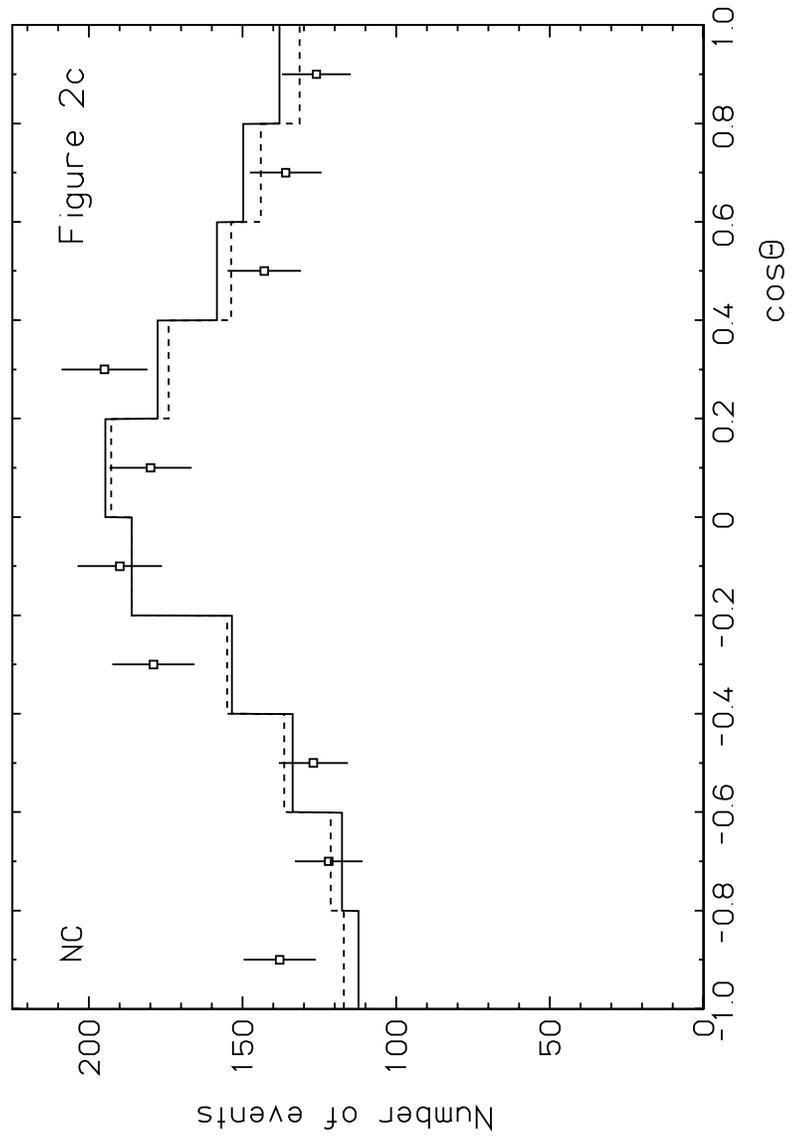
\end{document}